\documentclass[12pt,preprint]{aastex}

\usepackage{amsmath}

\begin{document}

\title{Angular Momentum Loss Mechanisms in Cataclysmic Variables below the Period Gap}
\author{
Yong Shao$^1$ and Xiang-Dong Li$^{1,2}$}

\affil{$^{1}$Department of Astronomy, Nanjing University, Nanjing 210093, China}

\affil{$^{2}$Key laboratory of Modern Astronomy and Astrophysics (Nanjing University), Ministry of
Education, Nanjing 210093, China}

\affil{$^{}$lixd@nju.edu.cn}

\begin{abstract}

Mass transfer in cataclysmic variables (CVs) is usually considered
to be caused by angular momentum loss (AML) driven by magnetic
braking and gravitational radiation (GR) above the period gap, and
solely by GR below the period gap. The best-fit revised model of CV
evolution recently by \citet{kbp11}, however, indicates that AML
rate below the period gap is $2.47(\pm 0.22)$ times the GR rate,
suggesting the existence of some other AML mechanisms. We consider
several kinds of consequential AML mechanisms often invoked in the
literature: isotropic wind from the accreting white dwarfs, outflows
from the Langrangian points, and the formation of a circumbinary
disk. We found that neither isotropic wind from the white dwarf nor
outflow from the $L_1$ point can explain the extra AML rate, while
ouflow from the $L_2$ point or a circumbinary disk can effectively
extract the angular momentum provided that $\sim (15-45)\%$ of the
transferred mass is lost from the binary. A more promising mechanism
is a circumbinary disk exerting gravitational torque on the binary.
In this case the mass loss fraction can be as low as $\lesssim
10^{-3}$.

\end{abstract}


\section{Introduction}
Cataclysmic variables (CVs) are short-period binaries consisting of
a white dwarf star accreting material from a lower-mass main-sequence
star that is overflowing its Roche-lobe \citep{w95}.
%
In the standard model, mass transfer in CVs is driven by angular momentum loss (AML) driven by
gravitational radiation (GR) \citep{kmg62} and magnetic braking (MB) \citep{vz81}.

The obital period ($P_{\rm orb}$) distribution of CVs has been summarized by \citet{rk03},
which is bimodal
with $\sim 45\%$ of CVs having period in the range of $\sim 3 - 16$ h, another $\sim 45\%$ with
$\sim 80$ min $- 2$ h, and the rest $\sim 10\%$ with $\sim 2-3$ h. The number of CVs in the period
interval $\sim 2-3$ h is small, and it is known as the period gap. In seeking a plausible
explanation of the period gap in CV evolution, several authors \citep{rvj83,sr83,lp94} proposed
the ``disrupted'' MB models
which are related to the transition of AML mechanisms. The general ideal is that, mass transfer
in CVs  above the period gap is primarily driven by MB at a rate rapid enough,
making the secondary star out of thermal equilibrium, so that the
secondary star is oversized and has larger radius than a main-sequence
star with the same mass. Along with mass transfer,  the secondary star loses mass gradually,
and finally become
fully convective when $P_{\rm orb}\simeq 3$ h. AML is now assumed to been caused solely by GR,
because MB is vanished.
Accordingly the oversized secondary star begins to
shrink and underfill its Roche-lobe in attempt to reach thermal equilibrium, cutting off matter transfer.
CVs becomes very faint, and virtually unobservable. When the Roche-lobe is filled again
due to orbit shrinking driven by GR
at $P_{\rm orb}\simeq 2$ h, mass transfer restarts from the secondary star.

Most recently \citet{kbp11} reconstructed the complete evolutionary path followed by
CVs, using  the observed mass-radius relationship of their secondary stars. For AML, they adopted
a scaled versions of the standard GR loss rate  and  the \citet{rvj83} MB law. With suitable
normalization parameters, $f_{GR}$ and $f_{MB}$, these recipes provide acceptable matches to the observed data. The best-fitting scaling factors are $f_{GR}=2.47\pm 0.22$ below  the gap,
and $f_{MB}=0.66\pm 0.05$ above, which describe the mass-radius data signicantly better than the standard model ($f_{GR}=f_{MB}=1$).

Here we focus on the origin of the enhanced AML below the gap, which
has already been mentioned before \citep[e.g.][]{kb99,p01,st01,bk03}. One obvious candidate is
the residual MB. Generally, the magnetic field of a low-mass
star is supposed to be anchored at the interface between the convective envelope and the
radiative core. For a fully convective star, such as the secondary star of CVs below the period gap,
the interface disappears, and MB is thought to be closed. However, there is strong evidence that fully convective stars are capable of generating significant magnetic fields \citep{spt00,aps03},
which might develop in a different way than low-mass main-sequence
stars \citep[for a discussion see \S8.5 in ][and references therein]{kbp11}.
At present it is not clear whether the
magnetically channelled stellar winds can produce MB strong enough to be consistent with
the model for CV evolution \citep[e.g.][]{lww94}.
In this paper we will examine other mechanisms that are needed to
account for the extra AML.



During the mass transfer processes, part of the transferred mass from the secondary star
may escape from the binary system,
carrying away the orbital angular momentum. AML associated with mass transfer is called
consequential angular momentum loss (CAML) \citep{kk95}. There are several types of CAML
due to different way of mass loss in the CV evolution
\citep{spv97}: (1) isotropic wind from the accreting white dwarf and surrounding accretion
disk \citep{kk95}, (2) outflow through the Lagrangian points $L_1$ or $L_{2}$
\citep{vvd98}, and (3) a circumbinary (CB) disk \citep{v94,ts01}.


The organization of this paper is as follows. We introduce the basics of orbital evolution
and possible CAML mechanisms  in Section 2.
In Section 3, we constrain possible AML machanisms to explain the extra AML rate
below the period gap. Discussions  and conclusions are presented in the final Section.

\section{Orbital evolution and CAML in CVs}

We first derive the relation between the CAML rate and the mass transfer rate $\dot{M}_{2}$, following
the approach in \citet{k88} and \citet{kbp11}.

We assume that the total
angular momentum of the binary is dominated by the
orbital angular momentum,
\begin{equation}
J=M_{1} M_{2} (\frac{Ga}{M})^{1/2},
\end{equation}
 where $a$ is the binary separation, $M_{1}$, $M_{2}$, and $M$ are the masses of the
 white dwarf, the secondary star,  and the binary system, respectively.   The relation between the binary separation $a$ and the orbital angular velocity
 $ \omega $  is given by the Kepler's third law,
\begin{equation}
GM=a^{3} \omega ^{2}.
\end{equation}
Logarithmic differentiation of Eq.~(1) gives
\begin{equation}
\frac{\dot{J}}{J} =\frac{\dot{M}_{1}}{M_{1}}+\frac{\dot{M}_{2}}{M_{2}}+\frac{\dot{a}}{2a}-\frac{\dot{M}}{2M}.
\end{equation}
The obital AML rate in CVs contains two items. One is the systemic AML due to MB and/or GR;
the other is CAML, which is related to the mass transfer rate. So the total obital AML rate can be writen as
\begin{equation}
\dot{J}=\dot{J}_{sys}+\dot{J}_{CAML}.
\end{equation}
When mass is transferred from the donor star to the white dwarf, we assume that a fraction $\delta $
of the matter flow escapes from the CV binary, i.e.,
\begin{equation}
\dot{M}=\delta\dot{M}_{2},
\end{equation}
which means that the actual accretion rate of the white dwarf is
\begin{equation}
\dot{M}_{1}=(\delta-1)\dot{M}_{2}.
\end{equation}
Considering that the CAML rate is related to the mass transfer rate $\dot{M}_{2}$,  we can write
\begin{equation}
\frac{\dot{J}_{CAML}}{J} =\nu\frac{\dot{M}_{2}}{M_{2}},
\end{equation}
where $\nu $ is a parameter as a function of $ \delta $ \citep{kbp11}.
In order to eliminate $ \dot{a}/a $ in Eq.~(3), we use the \citet{p71} formula for the Roche-lobe ridius
$R_{L}$ of the donor star,
\begin{equation}
R_{L}=0.462(M_{2}/M)^{1/3}a.
\end{equation}
Logarithmic differentiation of Eq.~(8) yields
\begin{equation}
\frac{\dot{R}_{L}}{R_{L}}=\frac{\dot{M}_{2}}{3M_{2}}-\frac{\dot{M}}{3M}+\frac{\dot{a}}{a}.
\end{equation}
For a Roche-lobe-filling donor star with steady mass transfer, variation of the stellar
radius $R_{2} $ and the Roche-lobe radius $ R_{L}$ should be in step, i.e.,
\begin{equation}
\frac{\dot{R}_{L}}{R_{L}} =\frac{\dot{R}_{2}}{R_{2}}.
\end{equation}
To deal with $\dot{R}_{2}$, we use the mass-radius relation$ R_{2} =M_{2} ^{\zeta}$ and
its logarithmic differentiation
\begin{equation}
\frac{\dot{R}_{2}}{R_{2}}=\zeta\frac{\dot{M}_{2}}{{M_{2}}},
\end{equation}
where $ \zeta $ is the mass-radius exponent. \citet{kbp11} showed
that after a CV donor emerges from the bottom of the period gap, the
exponent $\zeta$ evolves from $\simeq 0.8$ (in thermal equilibrium)
to $\simeq 1/3$ at the minimum period, and finally to $\simeq -1/3$
\citep[see also][]{rjw82}. They adopted broken-power-law fit to the
updated $M_2-R_2$ data in \citet{k06}, with $ \zeta=0.3 $ for $
M_{2} \la 0.07\,M_{\odot}$,  and $ \zeta=0.61 $ for $
0.07\,M_{\odot} \la M_{2} \la 0.2\,M_{\odot}$. Here $0.2 M_{\odot}$
and $0.07 M_{\odot}$ are the secondary masses just below the gap and
at the minimum period where the secondary stars become degenerate
and the orbital periods start to bounce back into the
period-increasing phase, respectively. Adopt $q=M_{2}/M_{1}$ and
typical white dwarf mass $M_{1}=0.6M_{\odot}$ in CVs, $ \zeta$
decreases from $\sim 0.61 $ to $\sim 0.3$ with decreasing $q$. 
In the following we calculate $\zeta$ by using the $M(R)$ relation 
derived from numerical calculations of CV binary evolution, which is
very close to the empirical one in \citet{kbp11}.

Combining Eqs.~(3)-(11), one can derive
\begin{equation}
\frac{\dot{M}_{2}}{M_{2}}=\frac{\dot{J}_{sys}}{JD},
\end{equation}
 where
\begin{equation}
D=(5/6+\zeta /2)-\frac{M_{2}}{M_{1}}+\delta(\frac{M_{2}}{M_{1}}-\frac{M_{2}}{3M})-\nu.
\end{equation}
Here the systemic AML mechanism below the gap is GR, and its rate is given by \citep{ll51}
\begin{equation}
\frac{\dot{J}_{GR}}{J}=-\frac{32}{5}\frac{G^{3}}{c^{5}}M_{1}M_{2}a^{-4}
\end{equation}

Next we consider the possible CAML mechanisms. For isotropic wind (e.g., outflows due to
nova explosions and/or wind emanating from the accretion disk), we assume that it
carries the white dwarf's
specific obital angular mometum  $ j_{1}$,
\begin{equation}
j_{1}=\frac{M_{2}}{M_{1}M}J.
\end{equation}
For outflows from the Lagrangian point $L_2$ the specific obital angular mometum is
\begin{equation}
j_{2}=a_{L_{2}}^{2}\omega,
\end{equation}
 where $a_{L_{2}}$ is the distance between the mass center of binary and the $L_{2}$ point.
Finally for CB disks we assume that it extracts the orbital angular momentum
from the binary by tidal force. The specific obital angular momentum $j_{3}$ is
 given by,
\begin{equation}
j_{3}=\gamma a^{2} \omega
\end{equation}
where $\gamma^2a$ is taken as the inner radius of the CB disk. Usually, $\gamma =1.5$ \citep{spv97}.

\section{Constraining the CAML machanisms below the period gap}

The best-fit revised model of CV evolution in \citet{kbp11} indicates that the AML rate below
the period gap is $2.47 \dot{J}_{GR}$. This means that some other AML
machanisms besides GR should work.
In the following we will discuss the feasibility of the AML related to isotropic wind, outflow
from $ L_{2} $ point, and CB disks, respectively.


\subsection{ Isotropic wind}
Combining Eqs.~(5), (7) and (15), we can derive the obital AML rate, $\dot{J}_{CAML, 1}$ of isotropic
wind as
\begin{equation}
\frac{\dot{J}_{CAML,1}}{J}=\delta\frac{M_{2}^{2}}{M_{1}M} \frac{\dot{M}_{2}}{M_{2}},
\end{equation}
and hence
\begin{equation}
\nu =\delta \frac{M_{2}^{2}}{M_{1}M}.
\end{equation}
Let $\dot{J}_{CAML,1}= 1.47\dot{J}_{GR}$
and  combined Eqs.~(12)-(14) and(18)-(20),  we have
\begin{equation}
\frac{2.47}{1.47}\frac{\delta M_{2}^{2}}{M_{1}M}=(5/6+\zeta /2)-\frac{M_{2}}{M_{1}}
+\delta(\frac{M_{2}}{M_{1}}-\frac{M_{2}}{3M}),
\end{equation}
which can be transformed into
\begin{equation}
\delta \simeq \frac{(2.5+1.5\zeta-3q)(1+q)}{2q(q-1)}.
\end{equation}
For CVs below the period gap, $ 0<q<0.33$, $\zeta > 0 $, so $ \delta $ is always negative,
indicating that the isotropic wind cannot provide the extra $1.47\dot{J}_{GR}$ AML.
This conclusion remains valid if we let most of the material to escape from the $L_1$ point
\citep{bk03}.

\subsection{Outflow from the $ L_{2}$ point }

Similar as in section 3.1, we derive the obital AML rate, $\dot{J}_{CAML,2}$,
of outflow from the $ L_{2}$ point by combining Eqs.~(5) and (16),
\begin{equation}
\frac{\dot{J}_{CAML,2}}{J}=\delta\frac{a_{L_{2}}^{2}}{a^{2}}\frac{M}{M_{1}}\frac{\dot{M}_{2}}{M_{2}},
\end{equation}
which gives
\begin{equation}
\nu =\delta \frac{a_{L_{2}}^{2}}{a^{2}}\frac{M}{M_{1}}.
\end{equation}
 Then we obtain the following relation
\begin{equation}
\frac{2.47}{1.47}\frac{\delta a_{L_{2}}^{2}M}{a^{2}M_{1}}=(5/6+\zeta /2)-\frac{M_{2}}{M_{1}}
+\delta(\frac{M_{2}}{M_{1}}-\frac{M_{2}}{3M}).
\end{equation}
After simplification we have
\begin{equation}
\delta\simeq \frac{(2.5+1.5\zeta-3q)(1+q)}{5(1+q)^2x^2-q(2+3q)},
\end{equation}
 where $ x=a_{L_{2}}/{a} $. For different $ q $, the values of $ x $ are given in \citet{m84}.
The relation between $ \delta $ and $ q $ is shown in Figure 1. It
is seen that the required value of $\delta$ ranges from $\sim 0.15$
to $\sim 0.45$.

\subsection{ CB disks}

The origin of the CB disks may stem from the  remnant of the late stage of
the common envelope evolution phase that formed the CVs, or matter outflow from CVs
during the mass transfer processes \citep{ts00}. Through tidal
interaction of the CB disk and CVs binary, the CB disk can extract obital angular momentun
from binary if part of the transferred mass flows into the disk rather onto the white dwarf.
The corresponding AML rate is,
\begin{equation}
\dot{J}_{CAML,3}=\delta\gamma\dot{M}_{2}a^{2}\omega.
\end{equation}
 Then
we can derive
\begin{equation}
\delta \simeq \frac{(2.5+1.5\zeta-3q)(1+q)}{7.56(1+q)^{2}-q(2+3q)}.
\end{equation}
The relation between $ \delta $ and $ q $ is shown in Figure 2. The
value of $\delta$ lies between $\sim 0.22$ and $\sim 0.4$,
comparable with that in the case of outflow from the $L_2$ point.

The above result is under the assumption that AML is only caused by
mass loss. Actually there is gravitational torque between the CB
disk after it is formed and the binary, which is more efficient to
extract angular momentum from the binary \citep{st01}. The AML rate
under this torque can be expressed as
\begin{equation}
\dot{J}_{CB}=\gamma(\frac{2\pi a^2}{P_{\rm orb}})\delta\dot{M}_{2}(\frac{t}{t_{vi}})^{1/3},
\end{equation}
where $t$ is the time since mass transfer begins, and
$t_{vi}$ is the viscous timescale at the inner edge of the CB disk, given by
$t_{vi}=2\gamma^3P_{\rm orb}/3\pi\alpha\beta^2$, where $\alpha$ is the viscosity
parameter \citep{ss73}, and $\beta$ is the ratio of the scale height to the radius of the disk.
Equation~(28) can be further simplified to be
\begin{equation}
\dot{J}_{CB}=A(GM)^{2/3}\delta\dot{M}_{2}t^{1/3},
\end{equation}
where $A=(3\alpha\beta^2/4)^{1/3}$. Similar as the derivation of
Eq.~(20), we obtain the following relation for the CB disk,
\begin{equation}
\frac{2.47}{1.47}\frac{AG^{1/6}M^{7/6}\delta t^{1/3}}{M_{1}a^{1/2}}=(5/6+\zeta /2)-\frac{M_{2}}{M_{1}}
+\delta(\frac{M_{2}}{M_{1}}-\frac{M_{2}}{3M}).
\end{equation}
Previous investigations \citep{ts01,t03,w05} suggest that very small values of $\delta (\ll 1)$
are required for CV evolution. Thus the third term on the rhs of Eq.~(30) can be neglected.
Considering the fact that $\zeta/2$ and $q$ also roughly counteract each other, and
$M\sim M_1$, Eq.~(30) is changed to be
\begin{eqnarray}
\delta &\simeq & \frac{1}{2A(2\pi)^{1/3}}(\frac{P_{\rm orb}}{t})^{1/3}        \nonumber    \\
             & \sim &  8\times 10^{-4} \alpha_{0.01}^{-1/3}\beta_{0.03}^{-2/3}
             (\frac{P_{\rm orb,2}}{t_{9}})^{1/3},
\end{eqnarray}
 where $\alpha_{0.01}=\alpha/0.01$, $\beta_{0.03}=\beta/0.03$ \citep{b04},
 $P_{\rm orb,2}=P_{\rm orb}/2$ hr, and $t_{9}=t/10^{9}$ yr. This is close to the result
 $\delta\sim 3\times 10^{-4}$ adopted in the numerical calculations by \citet{t03} for
 CV evolution.

To investigate the effect on AML of the uncertainties in treatment
of CB disks we have numerically solved Eq.~(30).  Figures 3 and 4
show the calculated values of $\delta$ as a function of $q$ for
different values of $\alpha$ and $\beta$. It is seen that generally
smaller $\alpha$ or $\beta$ corresponds to larger $\delta$. This is
easy to understand with Eq.~(29): smaller $\alpha$ or $\beta$
indicates less efficient angular momentum transfer within the disk,
which requires more mass input into the CB disk to guarantee enough
AML from the orbit.  Nevertheless, the values of $\delta$ are always
small ($<$ a few $10^{-3}$) when we change $\alpha$ from 0.001 to
0.1, and $\beta$ from 0.005 to 0.1. This implies that CB disks are
indeed very efficient in draining orbital angular momentum through
gravitational torques even with a very small mass input rate.

To show how CB disks can influence the evolution of CVs below the gap,
we have also performed binary evolution calculations
adopting an updated version of the stellar evolution code developed
by \citet{e71,e72} \citep[see also][]{hpe94,pteh95}. We set initial
solor chemical compositions (i.e., $X=0.7$, $Y=0.28$, and $Z=0.02$)
for the donor star, and take the ratio of mixing length to the
pressure scale height to be 2.0, and the convective overshooting
parameter  to be 0.12.

We follow the evolution of a CV just below the gap with a donor star
of mass $0.2 M_{\odot}$ and an orbital period $\sim 0.1$ d. For CB
disks we take $A\delta$ as one free parameter to assess its
influence on the evolution of CVs, since $\alpha$, $\beta$ and
$\delta$ are always combined together in evaluating the AML rate (see Eq.~[29]). For
typical values of $\alpha (=0.01)$ and $\beta (= 0.03)$ \citep[][and
references therein]{t03}, $A\simeq 0.02$. Considering the fact that
suitable value of $\delta$ may range from $\sim 10^{-7}$ to a few
$10^{-4}$ \citep{t03,w05}, we constrain the adopted value of
$A\delta$ to be less than  $\sim 10^{-5}$. Larger $A\delta$ may
cause unstable mass transfer. The stability of mass transfer can be
examined by comparing the Roche-lobe radius exponent $\zeta_{L}$ due
to mass loss with $\zeta$ \citep{spv97}, which are shown in  Figure 5 
as a function of $q$. From top to bottom, the solid curves represent
$\zeta_{L}$ with $A\delta$ ranging from $5\times10^{-5}$ to
$1\times10^{-5}$ in steps of $1\times10^{-5}$, and the dashed lines
show the mass-radius exponent $\zeta$ of 0.8, 1/3, and $-1/3$,
respectively. Since the mass transfer would be
unstable when $\zeta_{L}\geq\zeta$, we can derive that $A\delta$ should be
 $\lesssim (2-3)\times10^{-5}$ to guarantee stable mass transfer.

In Figure 6 we present examples of the evolutionary sequences of the
donor mass, orbital period, mass transfer rate, and the ratio of total 
AML rate and the AML rate due to GR ($N\equiv \dot{J}/\dot{J}_{GR}$) 
to show the influence CB disks. The panels from top to bottom
correspond to $A\delta=0$ (i.e., AML due to GR sololy),
$4\times10^{-6}$, $9\times10^{-6}$ and
$2\times10^{-5}$, respectively.
When
$A\delta=4\times10^{-6}$ (or $\delta=2\times 10^{-4}(A/0.02)^{-1}$),
the evolution seems similar to that with GR only. However, the mass transfer 
processes are actually accelerated, with $\sim 10\%$ higher average mass transfer 
rate and $\sim (10-15)\%$ higher AML rate than in the GR-only case. 
This tendency becomes more intense when $A\delta$
goes up. When $A\delta=2\times10^{-5}$, the average mass transfer rate is
enhanced by $\sim 70\%$, and the AML rate $\dot{J}$ becomes 
$\sim 2-3$ times
$\dot{J}_{GR}$. These results imply that a tiny fractional
input rate ($\delta\lesssim 10^{-3}$) into CB disks can
significantly change the evolution. This is in contrast with mass
loss-assissted AML mechanisms, which usually require a much larger
fraction of the transferred mass to leave the binary system
(although in real situation there might be multiple mechanisms work
simultaneously).

\section{Discussion and Conclusions}

The secular evolution of CVs is thought to be driven by the AML. Two mechanisms usually invoked
to account for the dissipation of AM are GR and MB of the secondary star.
However, this dual loss-mechanism cannot completely account for the magnitudes of
the mass-transfer rates inferred for some CVs, and for the large spread in the mass transfer rates
observed at a given orbital period \citep[e.g.][]{st01}. Likely additional AML mechanisms
include mass loss from the binary during the mass transfer process, which carries away the
orbital angular momentum from the binary.  Below the period gap in CVs evolution, AML
mechanism was usually considered to be driven by solely GR, while the best-fit result with
observations indicates that AML rate is about $ 2.47\dot{J}_{GR}$ \citep{kbp11}.
This offers a possibility to constrain the AML mechanisms besides GR. We consider
several kinds of CAML mechanisms often invoked in the literature: isotropic wind from the
accreting white dwarfs, outflows from the Langrangian points, and the formation of a CB disk.

We found that neither isotropic wind from the white dwarf nor outflow from the $L_1$ point
can explain  the extra $1.47 \dot{J}_{GR}$ AML rate, while outflow from the $ L_{2} $ point
or a CB disk is more  effective in extracting the  angular momentum. For a
$0.6 M_{\odot}$ white dwarf, the fraction $\delta$ of mass loss in the total transferred mass is
$\sim 0.15-0.45$ or $\sim 0.2-0.40 $, respectively. Actually it is found that $ \delta $ is allways less than
0.45 for different mass of the white dwarf in our calculations. Note that when mass is lost from
the $ L_{2} $ point, it is very likely to form a CB disk around the binary \citep{v94}, so
it is not surprised that the values of $\delta$ are very close in these two cases.

When the tidal interaction between the CB disk and the binary is included, the mass
transfer rate can be enhanced much more efficiently, and a very small fraction
($\delta\lesssim 10^{-3}$) of mass loss is required. This will suggest a much lighter CB disk
than in the former cases. The CB disks are thought to be large (up to several AU in radius)
and cool (a few thousand K at the inner edge to less than 1000 K at the outer edge),
with peak emission  in the infrared \citep{d02}. Although detection of excess infrared
emission from magnetic CVs provides observational support for the presence of cool gas
(and possibly a CB disk) surrounding CVs \citep{h06,d07,b07,h07}, there are still
many open questions associated with the formation of the CB disks. Future observations
with the measurement of the disk masses might distinguish the ways of angular momentum
transfer between the CB disk and the CV binary.

\begin{acknowledgements}
We are grateful to an anonymous referee for helpful comments.
This work was supported by the Natural Science Foundation of China
(under grant numbers 10873008 and 11133001), and the Ministry of Science and the
National Basic Research Program of China (973 Program 2009CB824800).

\end{acknowledgements}

\begin{figure}[h,t]

\centerline{\includegraphics[angle=0,width=0.40\textwidth]{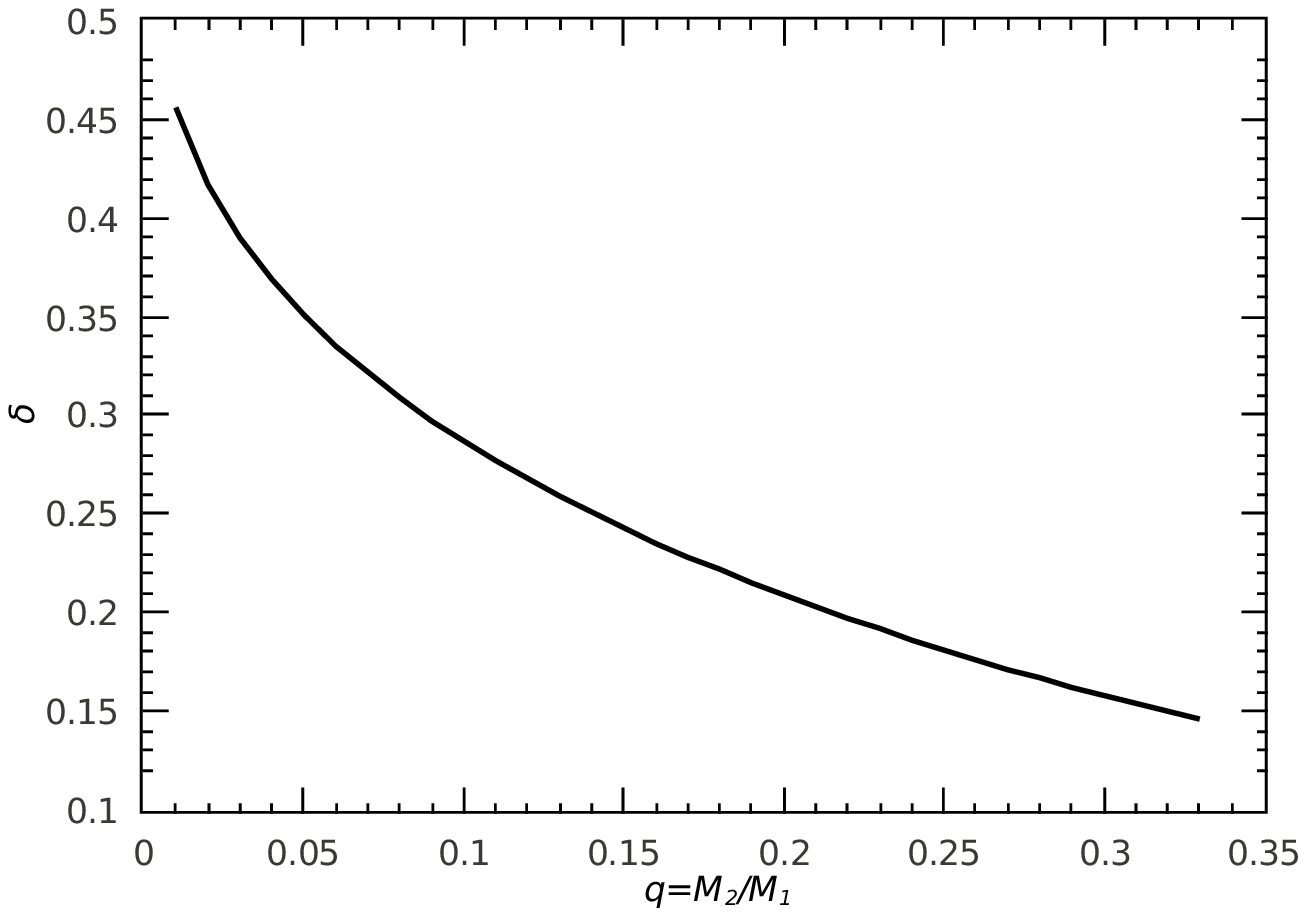}}
\caption{The relation between $ \delta $ and $ q$ in the case of outflow though the $L_{2} $ point.\label{figure1}}

\end{figure}

\begin{figure}[h,t]

\centerline{\includegraphics[angle=0,width=0.40\textwidth]{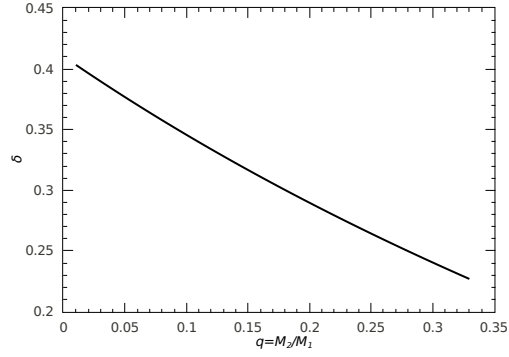}}
\caption{The relation between $ \delta $ and $ q$ in the case of CB disks.\label{figure2}}

\end{figure}

\begin{figure}[h,t]

\centerline{\includegraphics[angle=0,width=0.40\textwidth]{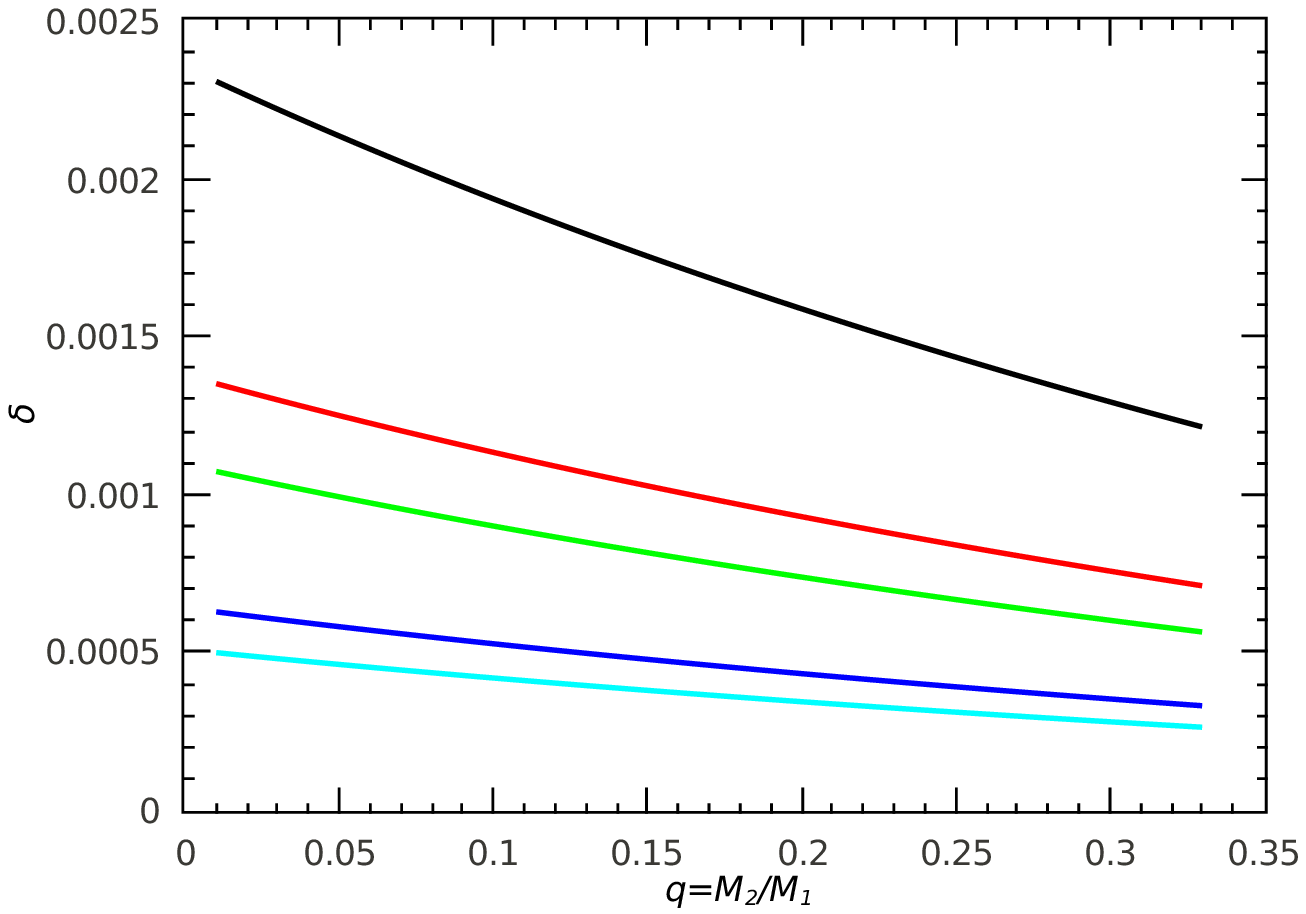}}
\caption{The fractional mass input rate, $ \delta $, as a function
of $q$ for CB disks with gravitational interaction with the binary.
Here we take $\beta=0.03$, and the curves from top to bottom
correspond to $\alpha=$  0.001, 0.005, 0.01, 0.05, and 0.1,
respectively. \label{figure3}}

\end{figure}

\begin{figure}[h,t]

\centerline{\includegraphics[angle=0,width=0.40\textwidth]{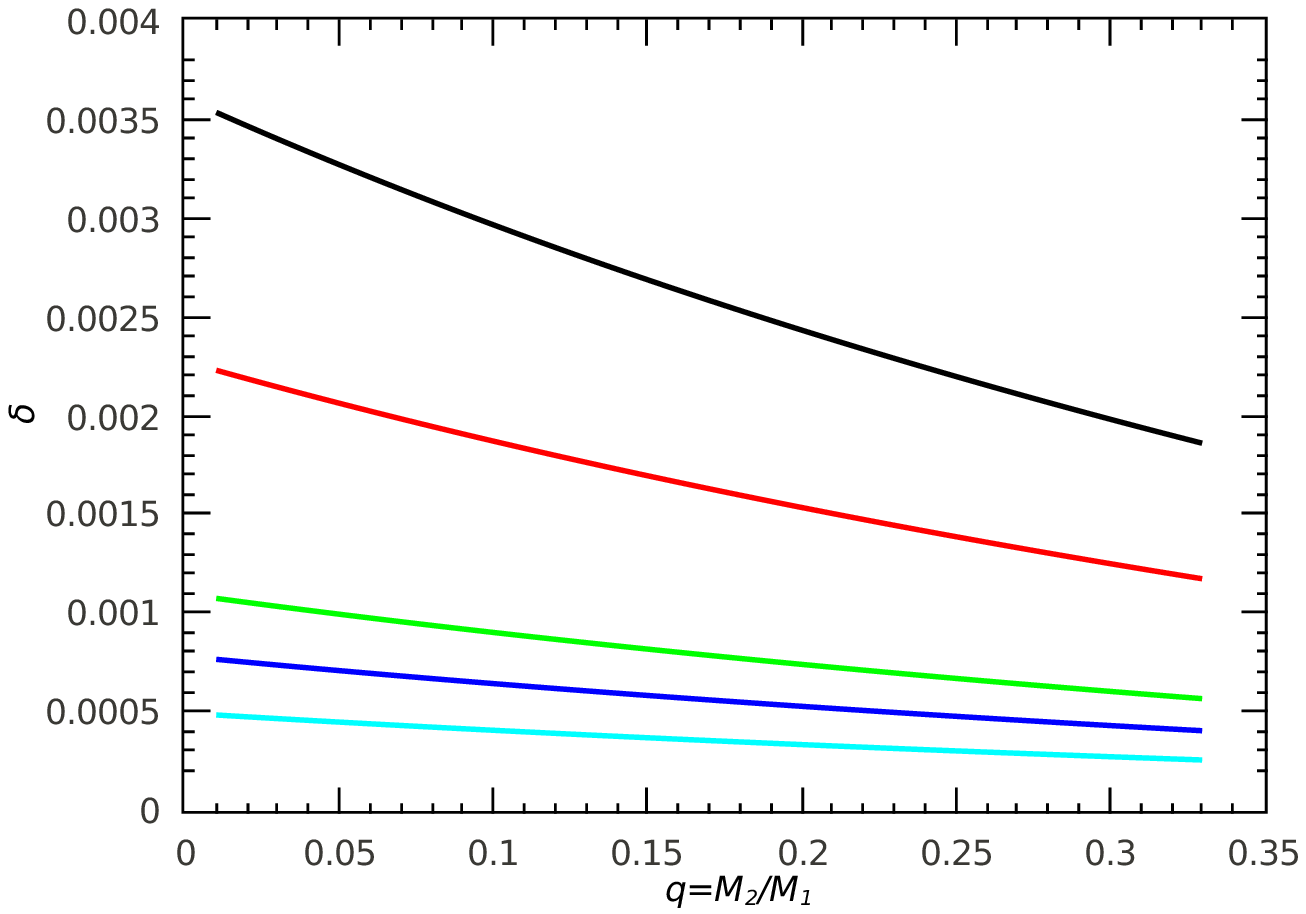}}
\caption{The fractional mass input rate, $ \delta $, as a function
of $q$ for CB disks with gravitational interaction with the binary.
Here we take $\alpha=0.01$, and the curves from top to bottom
correspond to $\beta=$  0.005, 0.01, 0.03, 0.05, and 0.1,
respectively. \label{figure4}}

\end{figure}

\begin{figure}[h,t]

\centerline{\includegraphics[angle=0,width=0.40\textwidth]{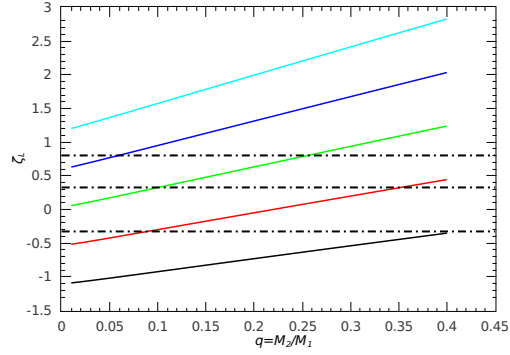}}
\caption{The Roche-lobe radius exponent $\zeta_{L}$ as a function of
$q$. The five solid curves represent $\zeta_{L}$ with different $A\delta$ from
$5\times10^{-5}$ to $1\times10^{-5}$ in steps of $1\times10^{-5}$
(from top to bottom). The dashed lines from top to bottom correspond
to the mass-radius exponent  $\zeta=0.8$, 1/3, and -1/3,
respectively.\label{figure6}}

\end{figure}

\begin{figure}[h,t]

\centerline{\includegraphics[angle=-90,width=1.00\textwidth]{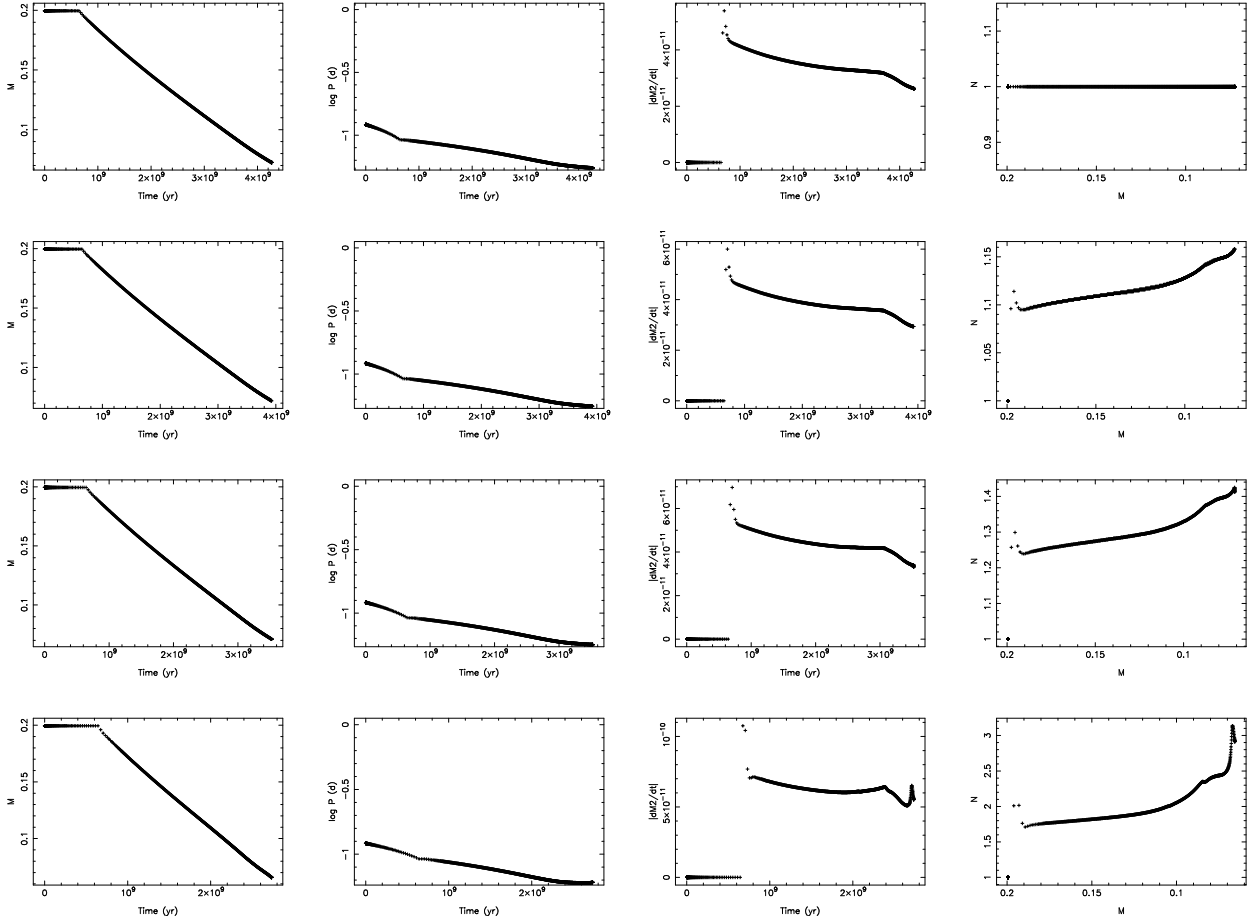}}
\caption{Evolution of the donor mass ($M_{\odot}$), orbital period
(d), mass transfer rate ($M_{\odot}$yr$^{-1}$), and the ratio ($N$) of total AML
rate and the AML rate due to GR.
The panels from top to bottom correspond to AML due to
GR sololy, and  both GR and CB disks with $A\delta= 4\times10^{-6}$, $ 9\times10^{-6}$,
and $2\times10^{-5}$, respectively.\label{figure5}}

\end{figure}

\clearpage

\label{lastpage}
\end{document}